\documentclass[aip,rsi,preprint,groupedaddress,showmsc,superscriptaddress,floatfix,pra,aps,showpacs,onecolumn]{revtex4}
\usepackage{amsmath,amsfonts,amssymb,caption,hyperref,color,epsfig,graphics,graphicx,latexsym,mathrsfs,revsymb,theorem,url,verbatim,epstopdf,cleveref}
\hypersetup{colorlinks,linkcolor={blue},citecolor={blue},urlcolor={red}}

\newenvironment{proof}{\noindent \textbf{{Proof.~} }}

\newtheorem{definition}{Definition}
\newtheorem{proposition}[definition]{Proposition}
\newtheorem{lemma}[definition]{Lemma}

\newtheorem{theorem}[definition]{Theorem}
\newtheorem{corollary}[definition]{Corollary}
\newtheorem{conjecture}[definition]{Conjecture}

\newtheorem{remark}[definition]{Remark}
\newtheorem{example}[definition]{Example}
\newtheorem{question}[definition]{Question}

\def\bcj{\begin{conjecture}}
\def\ecj{\end{conjecture}}
\def\bcr{\begin{corollary}}
\def\ecr{\end{corollary}}
\def\bd{\begin{definition}}
\def\ed{\end{definition}}
\def\bea{\begin{eqnarray}}
\def\eea{\end{eqnarray}}
\def\bem{\begin{enumerate}}
\def\eem{\end{enumerate}}
\def\bex{\begin{example}}
\def\eex{\end{example}}
\def\bim{\begin{itemize}}
\def\eim{\end{itemize}}
\def\bl{\begin{lemma}}
\def\el{\end{lemma}}
\def\bpf{\begin{proof}}
\def\epf{\end{proof}}
\def\bpp{\begin{proposition}}
\def\epp{\end{proposition}}
\def\bqu{\begin{question}}
\def\equ{\end{question}}
\def\br{\begin{remark}}
\def\er{\end{remark}}
\def\bt{\begin{theorem}}
\def\et{\end{theorem}}
\def\btb{\begin{tabular}}
\def\etb{\end{tabular}}

\def\tr{\mathop{\rm Tr}}
\begin{document}

\title{Entanglement criterion via general symmetric informationally complete measurement}

\author{Jun Li}
\thanks{Corresponding author: junlimath@buaa.edu.cn}
\affiliation{LMIB and School of Mathematical Sciences, Beihang University, Beijing 100191, China}
\author{Lin Chen}
\thanks{Corresponding author: linchen@buaa.edu.cn}
\affiliation{LMIB and School of Mathematical Sciences, Beihang University, Beijing 100191, China}
\affiliation{International Research Institute for Multidisciplinary Science, Beihang University, Beijing 100191, China}

\begin{abstract}
 We propose entanglement criteria for multipartite systems via symmetric informationally complete (SIC) measurement and general symmetric informationally complete (GSIC) measurement. We apply these criteria to detect entanglement of multipartite states, such as the convex of Bell states, entangled states mixed with white noise. It is shown that these criteria are stronger than some existing ones.
\end{abstract}
\maketitle

\section{INTRODUCTION}
\label{intro}
Due to the essential roles played in quantum information processing, entanglement \cite{MAN,RPMK,FMA,KSS,HPBO,JIV,CYSG} has been the subject of many studies in recent years. As one of the most distinctive features of quantum theory as compared to classical theory, entanglement has been widely used to quantum cryptography \cite{AKE}, teleportation \cite{chbg} and dense coding \cite{chbs}. Thus, the detection of entanglement of quantum states becomes important. Many criteria of detecting entanglement have been proposed. A well-known one is the positive partial transposition (PPT) criterion \cite{AP,mhph}, which is necessary and sufficient for the case of $m\times n$ quantum states with $mn\leq 6$. There are some other popular criteria, such as the realignment criterion \cite{OR}, the covariance matrix criterion \cite{ogph} and so on. Unlike the above criteria, the criteria of entanglement based on quantum measurements are more easily experimentally implemented. In \cite{csmh}, the criteria for two-qudit, multipartite, and continuous variable quantum states based on mutually unbiased bases (MUBs) have been presented. Then, the criteria based on MUBs were generalized to that based on mutually unbiased measurements, which can be always constructed with certain parameters for any $d$-dimensional space \cite{akgg,BCTM}.

Another kind of quantum measurements related with MUBs are known as symmetric informationally complete positive operator-valued measures (SIC-POVMs). More papers focus on SIC-POVMs with rank-one, which have been shown to exist in some certain dimensions. Rank-one SIC-POVMs are maximally efficient at determining the state of the system for quantum tomography \cite{JMRR}. The criteria via SIC-POVMs have been proposed for bipartite systems \cite{JWS}. Compared with SIC POVMs, the measurement operators of general SIC-POVMs (GSIC-POVMs) are not necessarily rank-one, which have been constructed in \cite{akgg2}. The criteria based on GSIC-POVMs have been proposed for bipartite systems \cite{LML,SSQ,BINC,YX}. As far as we are concerned, the entanglement criteria of multipartite systems based on the correlation matrices of SIC-POVMs and GSIC-POVMs have not been studied.

In this paper, we propose criteria of entanglement for tripartite systems via SIC-POVMs and GSIC-POVMs, which is a generalization of the criteria in \cite{JWS} and \cite{LML}. We also extend the criteria for tripartite system to multipartite system. First, we define the correlation matrices among SIC-POVMs for three subsystems for tripartite quantum states. Second, we apply the correlation matrices to the separable states and calculate the trace norm of the matrices. Then, the converse negative process of the entanglement criterion for tripartite system can be obtained. The same process can be extended to entanglement criteria based on GSIC-POVMs, which detect the entanglement of quantum states for arbitrary dimension. Third, we apply the criteria to detect entanglement of multipartite states, such as convex of Bell states, entangled states mixed with white noise and so on, and find that more entangled states can be detected. Moreover, our criterion is stronger than the existing one \cite{YAMA}.

This paper is organized as follows. In Sec. \ref{preli}, we review some concepts of SIC-POVMs and GSIC-POVMs and the related criteria for bipartite system. In Sec. \ref{result}, we provide the entanglement criteria based on SIC-POVMs and GSIC-POVMs and construct sufficient conditions in Theorems \ref{Theorem1} and \ref{Theorem2}. Moreover, the criteria are extended to multipartite system in Theorems \ref{Theorem4} and \ref{Theorem5}. In Sec. \ref{example}, we apply these criteria to several examples to verify its effectiveness. We conclude the paper in Sec. \ref{conclu}.

\section{PRELIMINARIES}
\label{preli}
In the $d$-dimensional Hilbert space, a POVM $M_{s}=\{\Pi _{k}\}_{k=1}^{d^{2}}$ with $d^{2}$ rank-1 operators acting on $\mathbb{C}^{d}$ is a SIC, if
\begin{eqnarray}\label{e1}
\Pi _{k}=\frac{1}{d}|\psi _{k}\rangle \langle\psi _{k}| ,
\end{eqnarray}
\begin{eqnarray}\label{e2}
\sum^{d^{2}} _{k=1}\Pi _{k}=\mathbb{I} ,
\end{eqnarray}
where $k=1, 2, \ldots , d^{2}$, $\mathbb{I}$ is the identity operator, and the vectors $|\psi _{k}\rangle$ satisfy
\begin{eqnarray}\label{e3}
|\langle\psi _{k}|\psi _{l}\rangle|^{2}=\frac{d\delta _{kl}+1}{d+1}, \quad k,l=1, 2, \ldots , d^{2}.
\end{eqnarray}
It has been conjectured that SIC POVMs exist in all finite dimensions \cite{GZA}. Analytical solutions have been found in dimensions $d=2-24, 28, 30, 31, 35, 37, 39, 43, 48, 124$, and numerical solutions have been found up to dimension $d=151$ \cite{CAF}. SIC-POVMs in low dimensions have been realized experimentally in various quantum systems \cite{NBE}.

For the $d$-dimensional Hilbert space, we renormalize the elements in the SIC-POVM,
\begin{eqnarray}\label{e4}
E_{k}\equiv \sqrt{\frac{d(d+1)}{2}}\Pi _{k}=\sqrt{\frac{d+1}{2d}}|\psi _{k}\rangle \langle\psi _{k}| .
\end{eqnarray}
For a bipartite state $\rho_{AB}$ acting on the Hilbert space $\rm H=\rm H^{A}\otimes  \rm H^{B}$ with the dimension $d=d^{A}\times d^{B}$, we assume that $\{E^{A}_{k}\}^{d^{2}_{A}}_{k=1}$ and $\{E^{B}_{k}\}^{d^{2}_{B}}_{k=1}$ be the renormalized SIC-POVMs of subsystems $A$ and $B$, respectively. For convenience, we abbreviate them as $E^{A}$ and $E^{B}$ .Denote
\begin{eqnarray}\label{e5}
[P_{s}]_{ij}:= \langle E^{A}_{i}\otimes E^{B}_{j}\rangle
\end{eqnarray}
as the linear correlations between $E^{A}$ and $E^{B}$.  We review an entanglement criterion for bipartite systems $AB$ via linear correlations of SIC.

\bl
\label{Lemma1}
\cite{JWS} For bipartite separable state $\rho_{AB}$, the following inequality holds,
\begin{eqnarray}\label{e6}
\lVert P_{s}\rVert_{tr}\leq 1,
\end{eqnarray}
where $\lVert P_{s}\rVert_{tr}=\tr \sqrt{P_{s}^{\dagger}P_{s}}$. The violation of inequality implies entanglement of $\rho_{AB}$.
\el

Unlike SIC, the general symmetric informationally complete (GSIC) measurement always exist for arbitrary dimension $d$. A POVM $\{M_{\alpha}\}_{\alpha=1}^{d^{2}}$ on $\mathbb{C}^{d}$ is said to be GSIC \cite{akgg2}, if

\begin{eqnarray}\label{g1}
\tr(M^{2}_{\alpha})=a, \quad \alpha=1, 2, \ldots, d^{2},
\end{eqnarray}
\begin{eqnarray}\label{g2}
\tr(M_{\alpha}M_{\beta})=\frac{1-da}{d(d^{2}-1)}, \quad 1\leq \alpha \neq \beta \leq d^{2},
\end{eqnarray}
where the parameter $a$ satisfies $\frac{1}{d^{3}}\leq a\leq \frac{1}{d^{2}}$. Here $a=\frac{1}{d^{2}}$ if and only if all $M_{\alpha}$ have rank one, which gives rise to a SIC-POVM \cite{JMRR}.

Consider two GSIC-POVMs $\{M_{\alpha}\}_{\alpha=1}^{d_{A}^{2}}$ with parameter $a_{A}$ and $\{M_{\alpha}\}_{\alpha=1}^{d_{B}^{2}}$ with parameter $a_{B}$ for subsystems $A$ and $B$, respectively. Denote
\begin{eqnarray}\label{g3}
[G_{s}]_{ij}:=\langle M^{A}_{i}\otimes M^{B}_{j}\rangle
\end{eqnarray}
as the linear correlations between $M^{A}$ and $M^{B}$. We review an entanglement criterion for bipartite systems $AB$ via linear correlations of GSIC.

\bl
\label{Lemma2}
\cite{LML} For bipartite separable state $\rho_{AB}$, the following inequality holds,
\begin{eqnarray}\label{gL}
\lVert G_{s}\rVert_{tr}\leq \sqrt{\frac{a_{A}d^{2}_{A}+1}{d_{A}(d_{A}+1)}}\sqrt{\frac{a_{B}d^{2}_{B}+1}{d_{B}(d_{B}+1)}},
\end{eqnarray}
where $\lVert G_{s}\rVert_{tr}=\tr \sqrt{G_{s}^{\dagger}G_{s}}$. The violation of inequality implies entanglement of $\rho_{AB}$.
\el

When $a_{R}=1/d_{R}^2$ for $R\in \{A, B\}$, Lemma \ref{Lemma2} is reduced to Lemma \ref{Lemma1}. The same method can be used to find the entanglement criterion of tripartite quantum states by constructing the correlation of SIC-POVMs or GSIC-POVMs. That is, Theorem \ref{Theorem1} and Theorem \ref{Theorem2}.

\section{ENTANGLEMENT CRITERION}
\label{result}
In this section, we investigate the generalization of the criteria in Lemmas \ref{Lemma1} and \ref{Lemma2} to tripartite and $N$-partite systems. Thus, we construct sufficient conditions for multipartite entanglement. The main results are Theorems \ref{Theorem1}, \ref{Theorem2},  \ref{Theorem4} and \ref{Theorem5}, where the first two are entanglement criteria for tripartite system and the last two are that for multipartite system.

\subsection{Entanglement criterion based on SIC-POVMs}
\label{sub:a}
For a tripartite state $\rho_{ABC}$ acting on the Hilbert sapce $\rm H=\rm H_{A}\otimes \rm H_{B}\otimes \rm H_{C}$ with dimension $d=d_{A}\times d_{B}\times d_{C}$, we consider the normalized SIC-POVMs $\{E^{A}_{i}\}^{d^{2}_{A}}_{i=1}$, $\{E^{B}_{j}\}^{d^{2}_{B}}_{j=1}$ and $\{E^{C}_{k}\}^{d^{2}_{C}}_{k=1}$ for the subsystems $A$, $B$ and $C$, respectively. We apply them to study the tripartite entanglement criterion. For any tripartite separable state
\begin{eqnarray}\label{e7}
\rho_{ABC}=\sum_{i}p_{i}\rho _{i}^{A}\otimes \rho _{i}^{B}\otimes \rho _{i}^{C},
\end{eqnarray}
the reduced states $\rho _{AB}$, $\rho _{AC}$ and $\rho _{BC}$ are also separable. Therefore, $\rho _{AB}$ satisfies the inequality (\ref{e6}) and similar statements hold for $\rho _{AC}$ and $\rho _{BC}$. That is
\begin{eqnarray}\label{e8}
\max\{\lVert P_{s}^{AB}\rVert_{tr}, \lVert P_{s}^{AC}\rVert_{tr}, \lVert P_{s}^{BC}\rVert_{tr}\}\leq 1,
\end{eqnarray}
where the correlation matrices $[P_{s}^{AB}]_{ij}=\langle E^{A}_{i}\otimes E^{B}_{j}\rangle _{\rho _{AB}}$, $[P_{s}^{AC}]_{ik}=\langle E^{A}_{i}\otimes E^{C}_{k}\rangle _{\rho _{AC}}$, $[P_{s}^{BC}]_{jk}=\langle E^{B}_{j}\otimes E^{C}_{k}\rangle _{\rho _{BC}}$. We extend the matrices to tripartite system.

Define the correlations among $E^{A}$, $E^{B}$ and $E^{C}$ as follows,
\begin{eqnarray}\label{e9}
[\mathcal{P}_{s}^{\underline{A}BC}]= \left(
\begin{array}{cccc}
P_{s1} & 0 & \cdots  & 0\\
0 & P_{s2} & \cdots & 0\\
\vdots & \vdots & \ddots &\vdots \\
0 & 0 & \cdots & P_{sd_{A}^{2}}
\end{array}
\right),
\end{eqnarray}
with $[P_{si}]_{jk}=\langle E^{A}_{i}\otimes E^{B}_{j}\otimes E^{C}_{k}\rangle$, $i=1, 2, \ldots, d_{A}^{2}$,
\begin{eqnarray}\label{e10}
[\mathcal{P}_{r}^{\underline{B}AC}]= \left(
\begin{array}{cccc}
P_{r1} & 0 & \cdots  & 0\\
0 & P_{r2} & \cdots & 0\\
\vdots & \vdots & \ddots &\vdots \\
0 & 0 & \cdots & P_{rd_{B}^{2}}
\end{array}
\right),
\end{eqnarray}
with $[P_{rj}]_{ik}=\langle E^{B}_{j}\otimes E^{A}_{i}\otimes E^{C}_{k}\rangle$, $j=1, 2, \ldots, d_{B}^{2}$,
\begin{eqnarray}\label{e11}
[\mathcal{P}_{t}^{\underline{C}AB}]= \left(
\begin{array}{cccc}
P_{t1} & 0 & \cdots  & 0\\
0 & P_{t2} & \cdots & 0\\
\vdots & \vdots & \ddots &\vdots \\
0 & 0 & \cdots & P_{td_{C}^{2}}
\end{array}
\right),
\end{eqnarray}
with $[P_{tk}]_{ij}=\langle E^{C}_{k}\otimes E^{A}_{i}\otimes E^{B}_{j}\rangle$, $k=1, 2, \ldots, d_{C}^{2}$. Now, we apply these correlations to the first main result in our paper.

\bt
\label{Theorem1}
For tripartite separable states (\ref{e7}), the inequality holds,
\begin{equation}
\max\{\lVert \mathcal{P}_{s}^{\underline{A}BC}\rVert_{tr}, \lVert \mathcal{P}_{r}^{\underline{B}AC}\rVert_{tr}, \lVert \mathcal{P}_{t}^{\underline{C}AB}\rVert_{tr}\}\leq 1.
\label{e12}
\end{equation}
\et

\bpf
For a product state $\rho =\rho _{A}\otimes \rho _{B}\otimes \rho _{C}$, we have
\begin{eqnarray}\label{e13}
\mathcal{P}_{s}^{\underline{A}BC}&&=\left(
\begin{array}{cccc}
e_{A,1} & 0 & \cdots  & 0\\
0 & e_{A,2} & \cdots & 0\\
\vdots & \vdots & \ddots &\vdots \\
0 & 0 & \cdots & e_{A,d^{2}_{A}}
\end{array}
\right)\otimes \nonumber\\
&&\left[\left(
\begin{array}{c}
e_{B,1}\\
e_{B,2}\\
\vdots \\
e_{B,d^{2}_{B}}
\end{array}
\right)
\left(
\begin{array}{cccc}
e_{C,1}&
e_{C,2}&
\cdots &
e_{C,d^{2}_{C}}
\end{array}
\right)\right],
\end{eqnarray}
where $e_{A,i}=\tr(\rho E^{A}_{i})$ for $i=1, 2, \ldots , d_{A}^{2}$, $e_{B,j}=\tr(\rho E^{B}_{j})$ for $j=1, 2, \ldots , d_{B}^{2}$ and $e_{C,k}=\tr(\rho E^{C}_{k})$ for $k=1, 2, \ldots , d_{C}^{2}$. Then
\begin{eqnarray}\label{e14}
\lVert \mathcal{P}_{s}^{\underline{A}BC}\rVert_{tr}
&&=\langle \textbf{e}_{A}|\textbf{e}_{A} \rangle^{\frac{1}{2}}\langle \textbf{e}_{B}|\textbf{e}_{B} \rangle^{\frac{1}{2}}\langle \textbf{e}_{C}|\textbf{e}_{C} \rangle^{\frac{1}{2}} \nonumber\\
&&=\sqrt{\sum _{i}e^{2}_{A,i}}\sqrt{\sum _{j}e^{2}_{B,j}}\sqrt{\sum _{k}e^{2}_{C,k}}\nonumber\\
&&\leq 1.
\end{eqnarray}
By employing the convexity property of the trace norm, we have $\lVert \mathcal{P}_{s}^{\underline{A}BC}\rVert_{tr}\leq 1$ for separable states in (\ref{e7}).
By changing the role of $\rho _{AB}$ with $\rho _{AC}$ and $\rho _{BC}$, respectively, we can obtain the inequality (\ref{e12}). \quad \quad $\square$
\epf

Thus, for the tripartite separable states, the inequalities (\ref{e8}) and (\ref{e12}) hold simultaneously. Violation of at least one of them for a tripartite state indicates that the state is entangled.

Next, we study the entanglement criterion for four-partite system based on SIC-POVMs. For a four-partite state $\rho_{ABCD}$ acting on the Hilbert sapce $\rm H=\rm H_{A}\otimes \rm H_{B}\otimes \rm H_{C}\otimes \rm H_{D}$ with dimension $d=d_{A}\times d_{B}\times d_{C}\times d_{D}$, we consider the normalized SIC-POVMs $\{E^{A}_{i}\}^{d^{2}_{A}}_{i=1}$, $\{E^{B}_{j}\}^{d^{2}_{B}}_{j=1}$, $\{E^{C}_{k}\}^{d^{2}_{C}}_{k=1}$and $\{E^{D}_{h}\}^{d^{2}_{D}}_{h=1}$ for the subsystems $A$, $B$, $C$ and $D$, respectively. Define the correlation matrices as follows,

\begin{eqnarray}\label{4c1}
\mathcal{P}_{s}^{AB|CD}=[\langle E_{i}\otimes E_{j}\rangle_{\rho_{AB}}]\otimes[\langle E_{k}\otimes E_{h}\rangle_{\rho_{CD}}],
\end{eqnarray}
with the reduced states $\rho _{AB}$ and $\rho _{CD}$. The correlation matrices $\mathcal{P}_{s}^{AC|BD}$ and $\mathcal{P}_{s}^{AD|BC}$ can be similarly defined. Next, we propose more correlation matrices,
\begin{eqnarray}\label{4c2}
\mathcal{P}_{s}^{A|BCD}=[\langle E_{i}\rangle_{\rho_{A}}]\otimes \mathcal{P}^{B|CD},
\end{eqnarray}
with the reduced states $\rho _{A}$, where $\mathcal{P}^{B|CD}$ is defined in (\ref{e9}). The correlation matrices $\mathcal{P}_{s}^{B|ACD}$, $\mathcal{P}_{s}^{C|ABD}$ and $\mathcal{P}_{s}^{D|ABC}$ can be similarly defined. Now we denote any one of these seven correlation matrices as $\mathcal{P}_{\rho_{ABCD}}$. The entanglement criterion for four-partite systems can be obtained.
\bcr
\label{corollary}
For four-partite separable states $\rho_{ABCD}=\sum_{i}p_{i}\rho _{i}^{A}\otimes \rho _{i}^{B}\otimes \rho _{i}^{C}\otimes \rho _{i}^{D}$, the inequality holds,
\begin{equation}
\lVert \mathcal{P}_{\rho_{ABCD}}\rVert_{tr}\leq 1.
\label{th3}
\end{equation}
\ecr

\bpf
For a product state $\rho =\rho _{A}\otimes \rho _{B}\otimes \rho _{C}\otimes \rho _{D}$, we have
\begin{eqnarray}\label{th31}
\mathcal{P}_{s}^{AB|CD}&&=\left[\left(
\begin{array}{c}
e_{A,1}\\
e_{A,2}\\
\vdots \\
e_{A,d^{2}_{A}}
\end{array}
\right)
\left(
\begin{array}{cccc}
e_{B,1}&
e_{B,2}&
\cdots &
e_{B,d^{2}_{B}}
\end{array}
\right)\right]\nonumber\\
&&\otimes \left[\left(
\begin{array}{c}
e_{C,1}\\
e_{C,2}\\
\vdots \\
e_{C,d^{2}_{C}}
\end{array}
\right)
\left(
\begin{array}{cccc}
e_{D,1}&
e_{D,2}&
\cdots &
e_{D,d^{2}_{D}}
\end{array}
\right)\right],
\end{eqnarray}
where $e_{A,i}=\tr(\rho E^{A}_{i})$ for $i=1, 2, \ldots , d_{A}^{2}$, $e_{B,j}=\tr(\rho E^{B}_{j})$ for $j=1, 2, \ldots , d_{B}^{2}$, $e_{C,k}=\tr(\rho E^{C}_{k})$ for $k=1, 2, \ldots , d_{C}^{2}$ and $e_{D,h}=\tr(\rho E^{D}_{h})$ for $h=1, 2, \ldots , d_{D}^{2}$. Then
\begin{eqnarray}\label{th32}
\lVert \mathcal{P}_{s}^{AB|CD}\rVert_{tr}
&&=\langle \textbf{e}_{A}|\textbf{e}_{A} \rangle^{\frac{1}{2}}\langle \textbf{e}_{B}|\textbf{e}_{B} \rangle^{\frac{1}{2}}\langle \textbf{e}_{C}|\textbf{e}_{C} \rangle^{\frac{1}{2}}\langle \textbf{e}_{D}|\textbf{e}_{D} \rangle^{\frac{1}{2}} \nonumber\\
&&=\sqrt{\sum _{i}e^{2}_{A,i}}\sqrt{\sum _{j}e^{2}_{B,j}}\sqrt{\sum _{k}e^{2}_{C,k}}\sqrt{\sum _{k}e^{2}_{D,h}}\nonumber\\
&&\leq 1.
\end{eqnarray}
By employing the convexity property of the trace norm, we have $\lVert \mathcal{P}_{s}^{AB|CD}\rVert_{tr}\leq 1$ for separable states $\rho_{ABCD}$. Similar calculation process shows that the trace norm of other correlation matrices is less than one. \quad \quad $\square$
\epf
The reduced states of separable state $\rho_{ABCD}$ are also separable. Therefore, they should satisfy the inequality (\ref{e6}) and (\ref{e12}). Obviously, for the four-partite separable states, the inequalities (\ref{e8}), (\ref{e12}) and (\ref{th3}) hold simultaneously. Violation of at least one of them for a four-partite state indicates that the state is entangled.

We extend the entanglement criterion to the N-partite system. For a N-partite state $\rho_{A_{1}A_{2}\ldots A_{N}}$ acting on the Hilbert sapce $\rm H=\rm H_{A_{1}}\otimes \rm H_{A_{2}}\otimes \ldots \otimes \rm H_{A_{N}}$ with dimension $d=d_{A_{1}}\times d_{A_{1}}\times \ldots\times d_{A_{N}}$, we consider the normalized SIC POVMs $\{E^{A_{i}}_{j}\}^{d^{2}_{A_{i}}}_{j=1}$ for the subsystem $A_{i}$, $i=1, 2, \ldots, N$. Define the correlation matrices as follows,
\begin{eqnarray}\label{nc1}
\mathcal{P}_{s}^{A_{1}|A_{2}\ldots A_{N}}=\langle E_{i}\rangle_{\rho_{A_{1}}}\otimes \mathcal{P}^{A_{2}|A_{3}\ldots A_{N}},
\end{eqnarray}
with the reduced states $\rho _{A_{1}}$, and $\mathcal{P}^{A_{2}|A_{3}\ldots A_{N}}$ is one of correlation matrices for $N-1$ partite systems. Next, we propose more correlation matrices,
\begin{eqnarray}\label{nc2}
\mathcal{P}_{s}^{A_{1}A_{2}|A_{3}\ldots A_{N}}=\langle E_{i}\otimes E_{j}\rangle_{\rho_{A_{1}A_{2}}}\otimes \mathcal{P}^{A_{3}|A_{4}\ldots A_{N}},
\end{eqnarray}
with the reduced states $\rho_{A_{1}A_{2}}$, and $\mathcal{P}^{A_{3}|A_{4}\ldots A_{N}}$ is one of correlation matrices for $N-2$ partite systems. The similar definition continues and the last kind of correlation matrix is defined by,
\begin{eqnarray}\label{nc3}
\mathcal{P}_{s}^{A_{1}A_{2}\ldots A_{\lfloor\frac{N}{2}\rfloor}|A_{\lfloor\frac{N}{2}\rfloor+1}\ldots A_{N}}=\mathcal{P}^{A_{1}|A_{2}\ldots A_{\lfloor\frac{N}{2}\rfloor}} \otimes \mathcal{P}^{A_{\lfloor\frac{N}{2}\rfloor+1}|A_{\lfloor\frac{N}{2}\rfloor+2}\ldots A_{N}},
\end{eqnarray}
where $\mathcal{P}^{A_{1}|A_{2}\ldots A_{\lfloor\frac{N}{2}\rfloor}}$ is one of correlation matrices for $\lfloor\frac{N}{2}\rfloor$ partite systems and $\mathcal{P}^{A_{\lfloor\frac{N}{2}\rfloor+1}|A_{\lfloor\frac{N}{2}\rfloor+2}\ldots A_{N}}$ is one of correlation matrices for $N-\lfloor\frac{N}{2}\rfloor$ partite systems. Now we denote any one of these correlation matrices as $\mathcal{P}_{\rho_{A_{1}A_{2}\ldots A_{N}}}$. The entanglement criterion for $N$-partite systems can be obtained.
\bt
\label{Theorem4}
For $N$-partite separable states $\rho_{A_{1}A_{2}\ldots A_{N}}=\sum_{i}p_{i}\rho _{i}^{A_{1}}\otimes \rho _{i}^{A_{2}}\otimes \ldots\otimes \rho _{i}^{A_{N}}$, the inequality holds,
\begin{equation}
\lVert \mathcal{P}_{\rho_{A_{1}A_{2}\ldots A_{N}}}\rVert_{tr}\leq 1.
\label{th4}
\end{equation}
\et
The proof of inequality (\ref{th4}) follows from the proof of inequality (\ref{th3}).
\subsection{Entanglement criterion based on GSIC-POVMs}
There is no proof that SIC-POVMs always exist in any dimension \cite{GZA}. Therefore, the entanglement criterion based on SIC-POVMs can not be applied to high dimensional quantum system. Thus, we extend the entanglement criterion in section \ref{sub:a} to the criterion based on GSIC-POVMs. In this way, we can detect the entanglement of quantum states for arbitrary dimension. We first study the entanglement criterion based on GSIC-POVMs for tripartite system. Consider GSIC-POVMs $\{M^{A}_{i}\}^{d^{2}_{A}}_{i=1}$ with parameter $a_{A}$, $\{M^{B}_{j}\}^{d^{2}_{B}}_{j=1}$ with parameter $a_{B}$ and $\{M^{C}_{k}\}^{d^{2}_{C}}_{k=1}$ with parameter $a_{C}$ for the subsystems $A$, $B$ and $C$, respectively.

For the tripartite separable state (\ref{e7}), the reduced states $\rho _{AB}$, $\rho _{AC}$ and $\rho _{BC}$ are also separable. Therefore, $\rho _{AB}$ satisfies the inequality (\ref{gL}) and similar statements hold for $\rho _{AC}$ and $\rho _{BC}$, that is

\begin{eqnarray}\label{gabc}
&&\lVert G_{s}^{AB}\rVert_{tr}\leq \sqrt{\frac{a_{A}d^{2}_{A}+1}{d_{A}(d_{A}+1)}}\sqrt{\frac{a_{B}d^{2}_{B}+1}{d_{B}(d_{B}+1)}}, \nonumber\\
&&\lVert G_{s}^{AC}\rVert_{tr}\leq \sqrt{\frac{a_{A}d^{2}_{A}+1}{d_{A}(d_{A}+1)}}\sqrt{\frac{a_{C}d^{2}_{C}+1}{d_{C}(d_{C}+1)}},\nonumber\\
&&\lVert G_{s}^{BC}\rVert_{tr}\leq \sqrt{\frac{a_{B}d^{2}_{B}+1}{d_{B}(d_{B}+1)}}\sqrt{\frac{a_{C}d^{2}_{C}+1}{d_{C}(d_{C}+1)}},
\end{eqnarray}
where the correlation matrices $[G_{s}^{AB}]_{ij}=\langle M^{A}_{i}\otimes M^{B}_{j}\rangle _{\rho _{AB}}$, $[G_{s}^{AC}]_{ik}=\langle M^{A}_{i}\otimes M^{C}_{k}\rangle _{\rho _{AC}}$, $[G_{s}^{BC}]_{jk}=\langle M^{B}_{j}\otimes M^{C}_{k}\rangle _{\rho _{BC}}$. We extend them to tripartite system.

Define the same correlations of GSIC-POVMs $\mathcal{G}_{s}^{\underline{A}BC}$, $\mathcal{G}_{r}^{\underline{B}AC}$, and $\mathcal{G}_{t}^{\underline{C}AB}$ among $M^{A}$, $M^{B}$ and $M^{C}$ as (\ref{e9}), (\ref{e10}) and (\ref{e11}). We apply these correlations to the second main result in our paper.

\bt
\label{Theorem2}
For tripartite separable states (\ref{e7}), the inequality holds,
\begin{eqnarray}\label{gt}
&&\max\{\lVert \mathcal{G}_{s}^{\underline{A}BC}\rVert_{tr},\lVert \mathcal{G}_{r}^{\underline{B}AC}\rVert_{tr},\lVert \mathcal{G}_{t}^{\underline{C}AB}\rVert_{tr}\}\nonumber\\
&&\leq\sqrt{\frac{a_{A}d^{2}_{A}+1}{d_{A}(d_{A}+1)}}\sqrt{\frac{a_{B}d^{2}_{B}+1}{d_{B}(d_{B}+1)}}\sqrt{\frac{a_{C}d^{2}_{C}+1}{d_{C}(d_{C}+1)}}.
\end{eqnarray}
\et

\bpf
For a product state $\rho =\rho _{A}\otimes \rho _{B}\otimes \rho _{C}$, we have
\begin{eqnarray}\label{th20}
&&\mathcal{G}_{s}^{\underline{A}BC}=\left(
\begin{array}{cccc}
m_{A,1} & 0 & \cdots  & 0\\
0 & m_{A,2} & \cdots & 0\\
\vdots & \vdots & \ddots &\vdots \\
0 & 0 & \cdots & m_{A,d^{2}_{A}}
\end{array}
\right)\otimes \nonumber\\
&&\left[\left(
\begin{array}{c}
m_{B,1}\\
m_{B,2}\\
\vdots \\
m_{B,d^{2}_{B}}
\end{array}
\right)
\left(
\begin{array}{cccc}
m_{C,1}&
m_{C,2}&
\cdots &
m_{C,d^{2}_{C}}
\end{array}
\right)\right],
\end{eqnarray}
where $m_{A,i}=\tr(\rho M^{A}_{i})$ for $i=1, 2, \ldots , d_{A}^{2}$, $m_{B,j}=\tr(\rho M^{B}_{j})$ for $j=1, 2, \ldots , d_{B}^{2}$ and $m_{C,k}=\tr(\rho M^{C}_{k})$ for $k=1, 2, \ldots , d_{C}^{2}$. Then
\begin{eqnarray}\label{th21}
\lVert \mathcal{G}_{s}^{\underline{A}BC}\rVert_{tr}&&=\langle \textbf{m}_{A}|\textbf{m}_{A} \rangle^{\frac{1}{2}}\langle \textbf{m}_{B}|\textbf{m}_{B} \rangle^{\frac{1}{2}}\langle \textbf{m}_{C}|\textbf{m}_{C} \rangle^{\frac{1}{2}} \nonumber\\
&&=\sqrt{\sum _{i}m^{2}_{A,i}}\sqrt{\sum _{j}m^{2}_{B,j}}\sqrt{\sum _{k}m^{2}_{C,k}}\nonumber\\
&&\leq \sqrt{\frac{a_{A}d^{2}_{A}+1}{d_{A}(d_{A}+1)}}\sqrt{\frac{a_{B}d^{2}_{B}+1}{d_{B}(d_{B}+1)}}\sqrt{\frac{a_{C}d^{2}_{C}+1}{d_{C}(d_{C}+1)}}.
\end{eqnarray}
By employing the convexity property of the trace norm, we have $\lVert \mathcal{G}_{s}^{\underline{A}BC}\rVert_{tr}\leq \sqrt{\frac{a_{A}d^{2}_{A}+1}{d_{A}(d_{A}+1)}}\sqrt{\frac{a_{B}d^{2}_{B}+1}{d_{B}(d_{B}+1)}}\sqrt{\frac{a_{C}d^{2}_{C}+1}{d_{C}(d_{C}+1)}}$ for separable states in (\ref{e7}).
By changing the role of $\rho _{AB}$ with $\rho _{AC}$ and $\rho _{BC}$, respectively, we can also obtain the two other inequalities of (\ref{gt}). \quad \quad $\square$
\epf

Obviously, for the tripartite separable states, the inequalities (\ref{gabc}) and (\ref{gt}) hold simultaneously. Violation of at least one of them for a tripartite state indicates that the state is entangled.

We explain the connection between Theorems \ref{Theorem1} and \ref{Theorem2} as follows. For $d$-dimension Hilbert space, we renormalize the elements in the GSIC-POVM,
\begin{eqnarray}\label{g4}
M'_{k}\equiv \sqrt{\frac{d(d+1)}{2}}M_{k}.
\end{eqnarray}
The correlations of renormalized GSIC-POVM among $M'^{A}$, $M'^{B}$ and $M'^{C}$ are marked by ${\mathcal{G}'}_{s}^{\underline{A}BC}$, ${\mathcal{G}'}_{r}^{\underline{B}AC}$, and ${\mathcal{G}'}_{t}^{\underline{C}AB}$. We show the following criterion.

\bcr
Take $a_{R}=\frac{1}{d_{R}^{2}}$, $R\in \{A, B, C\}$. For tripartite separable states (\ref{e7}),
\begin{eqnarray}\label{cr1}
\max\{\lVert {\mathcal{G}'}_{s}^{\underline{A}BC}\rVert_{tr}, \lVert {\mathcal{G}'}_{r}^{\underline{B}AC}\rVert_{tr}, \lVert {\mathcal{G}'}_{t}^{\underline{C}AB}\rVert_{tr}\}\leq 1.
\end{eqnarray}
The criterion is reduced to the criterion based on SIC-POVMs in Theorem \ref{Theorem1}.
\ecr

Next, we extend the tripartite entanglement criterion in Theorem \ref{Theorem2} to $N$-partite system. Consider the GSIC-POVMs $\{M^{A_{i}}_{j}\}^{d^{2}_{A_{i}}}_{j=1}$ with parameter $a_{A_{i}}$ for the subsystem $A_{i}$, $i=1, 2, \ldots, N$. The correlation matrices for $N$-partite systems based on GSIC-POVM can be defined, which are similar with (\ref{nc1}), (\ref{nc2}) and (\ref{nc3}). Entanglement criterion based on GSIC-POVM can be obtained,
\bt
\label{Theorem5}
For $N$-partite separable states $\rho_{A_{1}A_{2}\ldots A_{N}}=\sum_{i}p_{i}\rho _{i}^{A_{1}}\otimes \rho _{i}^{A_{2}}\otimes \ldots\otimes \rho _{i}^{A_{N}}$, the inequality holds,
\begin{equation}
\lVert \mathcal{G}_{\rho_{A_{1}A_{2}\ldots A_{N}}}\rVert_{tr}\leq \Pi_{i=1}^{N}\sqrt{\frac{a_{A_{i}}d^{2}_{A_{i}}+1}{d_{A_{i}}(d_{A_{i}}+1)}}.
\label{th5}
\end{equation}
\et
The proof of the inequalities (\ref{th5}) follows from the proof in (\ref{th3}). Take $a_{R}=\frac{1}{d_{R}^{2}}$, $R\in \{A_{1}, A_{2}, \ldots A_{N}\}$, the criterion in Theorem \ref{Theorem5} is reduced to the criterion based on SIC-POVMs in Theorem \ref{Theorem4}.
\section{EXAMPLES}
\label{example}
In order to show the effectiveness of the entanglement criteria, we consider some examples in this section.

In the $d$-dimensional Hilbert space, there are infinite SIC-POVMs for some specific $d$ \cite{GZA}. Here, we list a set of SIC-POVM vectors for $d = 2$ \cite{GMC}.
\begin{eqnarray}\label{e15}
&&|\varphi _{1}\rangle=\left(
\begin{array}{c}
1\\
0
\end{array}\right), \quad  \quad \quad\quad
|\varphi _{2}\rangle=\frac{1}{\sqrt{3}}\left(
\begin{array}{c}
1\\
\sqrt{2}
\end{array}\right),\nonumber\\
&&|\varphi _{3}\rangle=\frac{i}{\sqrt{3}}\left(
\begin{array}{c}
e^{\frac{-i\pi }{3}}\\
-\sqrt{2}
\end{array}\right),\quad
|\varphi _{4}\rangle=\frac{1}{\sqrt{3}}\left(
\begin{array}{c}
1\\
-\sqrt{2}e^{\frac{-i\pi}{3}}
\end{array}\right).
\end{eqnarray}

Let $\{M_{\alpha}\}_{\alpha=1}^{d^{2}}$ be a set of GSIC POVM on $\mathbb{C}^{d}$ with parameter $a$, and $\{\bar{M}_{\alpha}\}_{\alpha=1}^{d^{2}}$ a set of GSIC POVM with the same parameter $a$, where $\bar{M}_{\alpha}$ is the conjugation of $M_{\alpha}$.

For any non-zero $t\in[-0.068, 0.068]$, the four matrices form a GSIC-POVM for $d=2$,
\begin{eqnarray}\label{GSIC1}
&&M_{\alpha}=\frac{1}{4}I+t(F_{4}-6F_{\alpha}) \quad for \quad \alpha=1, 2, 3,\nonumber\\
&&M_{4}=\frac{1}{4}I+3tF_{4},
\end{eqnarray}
where
\begin{eqnarray*}
&&F_{1}=\frac{1}{\sqrt{2}}\left(
  \begin{array}{cc}
    0 & 1 \\
    1 & 0 \\
  \end{array}
\right),\quad \quad
F_{2}=\frac{1}{\sqrt{2}}\left(
                          \begin{array}{cc}
                            0 & i \\
                            -i & 0 \\
                          \end{array}
                        \right),\nonumber\\
&&F_{3}=\frac{1}{\sqrt{2}}\left(
                            \begin{array}{cc}
                              1 & 0 \\
                              0 & -1 \\
                            \end{array}
                          \right),\quad \quad
F_{4}=\frac{1}{\sqrt{2}}\left(
                          \begin{array}{cc}
                            1 & 1+i \\
                            1-i & 1 \\
                          \end{array}
                        \right).
\end{eqnarray*}

For any non-zero $t\in[-0.012, 0.012]$, the nine matrices form a GSIC-POVM for $d=3$,
\begin{eqnarray}\label{GSIC2}
&&M_{\alpha}=\frac{1}{9}I+t(G_{9}-12G_{\alpha}) \quad for \quad \alpha=1, 2, \ldots, 8,\nonumber\\
&&M_{9}=\frac{1}{9}I+4tG_{9},
\end{eqnarray}
where
\begin{eqnarray*}
&&G_{1}=\left(
          \begin{array}{ccc}
            \frac{1}{\sqrt{2}} & 0 & 0 \\
            0 & -\frac{1}{\sqrt{2}} & 0 \\
            0 & 0 & 0 \\
          \end{array}
        \right),\quad \quad
G_{2}=\left(
        \begin{array}{ccc}
          0 & \frac{1}{\sqrt{2}} & 0 \\
          \frac{1}{\sqrt{2}} & 0 & 0 \\
          0 & 0 & 0 \\
        \end{array}
      \right),\quad \quad
G_{3}=\left(
        \begin{array}{ccc}
          0 & 0 & \frac{1}{\sqrt{2}} \\
          0 & 0 & 0 \\
          \frac{1}{\sqrt{2}} & 0 & 0 \\
        \end{array}
      \right),\nonumber\\
&&G_{4}=\left(
          \begin{array}{ccc}
            0 & -\frac{i}{\sqrt{2}} & 0 \\
            \frac{i}{\sqrt{2}} & 0 & 0 \\
            0 & 0 & 0 \\
          \end{array}
        \right),\quad \quad
G_{5}=\left(
        \begin{array}{ccc}
          \frac{1}{\sqrt{6}} & 0 & 0 \\
          0 & \frac{1}{\sqrt{6}} & 0 \\
          0 & 0 & -\sqrt{\frac{2}{3}} \\
        \end{array}
      \right),\quad \quad
G_{6}=\left(
        \begin{array}{ccc}
          0 & 0 & 0 \\
          0 & 0 & \frac{1}{\sqrt{2}} \\
          0 & \frac{1}{\sqrt{2}} & 0 \\
        \end{array}
      \right),\nonumber\\
&&G_{7}=\left(
          \begin{array}{ccc}
            0 & 0 & -\frac{i}{\sqrt{2}} \\
            0 & 0 & 0 \\
            \frac{i}{\sqrt{2}} & 0 & 0 \\
          \end{array}
        \right),\quad \quad
G_{8}=\left(
        \begin{array}{ccc}
          0 & 0 & 0 \\
          0 & 0 & -\frac{i}{\sqrt{2}} \\
          0 & \frac{i}{\sqrt{2}} & 0 \\
        \end{array}
      \right),\quad \quad
G_{9}=\left(
        \begin{array}{ccc}
          \frac{1}{\sqrt{2}}+\frac{1}{\sqrt{6}} & \frac{1-i}{\sqrt{2}} & \frac{1-i}{\sqrt{2}} \\
          \frac{1+i}{\sqrt{2}} & -\frac{1}{\sqrt{2}}+\frac{1}{\sqrt{6}} & \frac{1-i}{\sqrt{2}} \\
          \frac{1+i}{\sqrt{2}} & \frac{1+i}{\sqrt{2}} & -\sqrt{\frac{2}{3}} \\
        \end{array}
      \right).
\end{eqnarray*}
We apply the two sets of GSIC-POVMs $\{M_{\alpha}\}_{\alpha=1}^{d^{2}}$ and $\{\bar{M}_{\alpha}\}_{\alpha=1}^{d^{2}}$ to Theorem \ref{Theorem2} for entanglement detecting.

\bex
\label{example1}
We consider the three-qubit state
\begin{eqnarray*}
\varrho&&=\frac{1}{3}(|\psi ^{+}\rangle \langle\psi ^{+}|_{AB}\otimes |0\rangle \langle 0|_{C}+|\psi ^{+}\rangle \langle\psi ^{+}|_{AC}\otimes |0\rangle \langle 0|_{B}\nonumber\\
&&+|0\rangle \langle 0|_{A}\otimes |\psi ^{+}\rangle \langle\psi ^{+}|_{BC}),\nonumber
\end{eqnarray*}
with $|\psi ^{+}\rangle=\frac{1}{\sqrt{2}}(|00\rangle +|11\rangle)$. The correlation matrix is calculated by SIC-POVM vectors in (\ref{e15}),
\begin{eqnarray}
\mathcal{P}_{\varrho}^{\underline{A}BC}=\left(
\begin{array}{cccc}
\frac{\sqrt{3}}{3}& 0 & 0 & 0\\
0 & \frac{2\sqrt{3}}{9} & 0 & 0\\
0 & 0 & \frac{2\sqrt{3}}{9} & 0 \\
0 & 0 & 0 & \frac{2\sqrt{3}}{9}
\end{array}
\right)\otimes \left(
\begin{array}{cccc}
\frac{3}{8} & \frac{5}{24} & \frac{5}{24} & \frac{5}{24}\\
\frac{5}{24} & \frac{5}{24} & \frac{1}{8} & \frac{1}{8}\\
\frac{5}{24} & \frac{1}{8} & \frac{1}{8} & \frac{5}{24} \\
\frac{5}{24} & \frac{1}{8} & \frac{5}{24} & \frac{1}{8}
\end{array}
\right),\nonumber
\end{eqnarray}
we have
$\parallel\mathcal{P}_{\varrho}^{\underline{A}BC}\parallel_{tr}=2.687> 1.$ Thus, Theorem \ref{Theorem1} can detect the entanglement of $\varrho$. It coincides with the result in \cite{ZMJ}.
\begin{figure}[h]
\centering
\includegraphics[width=9in]{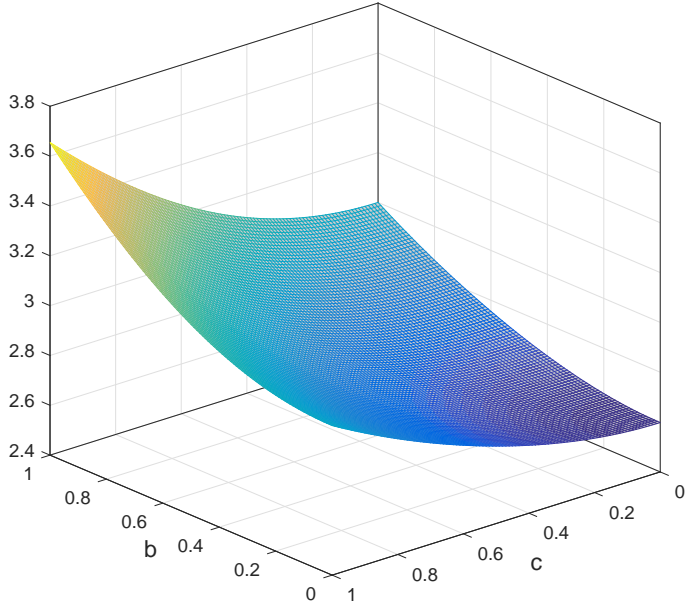}
\caption{The value of $\parallel\mathcal{P}_{\varrho^{'}}^{\underline{A}BC}\parallel_{tr}$ for $b\in[0, 1]$, $c\in[0, 1]$.}
\label{fig:ex1abc}
\end{figure}

Now, we consider the more general quantum state,
\begin{eqnarray*}
\varrho^{'}&&=a|\psi ^{+}\rangle \langle\psi ^{+}|_{AB}\otimes |0\rangle \langle 0|_{C}+b|\psi ^{+}\rangle \langle\psi ^{+}|_{AC}\otimes |0\rangle \langle 0|_{B}\nonumber\\
&&+c|0\rangle \langle 0|_{A}\otimes |\psi ^{+}\rangle \langle\psi ^{+}|_{BC},\nonumber
\end{eqnarray*}
with $a+b+c=1$, $a, b, c\geq 0$.
The correlation matrix can be obtained as
\begin{eqnarray}
\mathcal{P}_{\varrho^{'}}^{\underline{A}BC}&&=\left(
\begin{array}{cccc}
\frac{\sqrt{3}}{4}(1+c)& 0 & 0 & 0\\
0 & \frac{\sqrt{3}}{6}(a+b)+\frac{\sqrt{3}}{12}(1+c) & 0 & 0\\
0 & 0 & \frac{\sqrt{3}}{6}(a+b)+\frac{\sqrt{3}}{12}(1+c) & 0 \\
0 & 0 & 0 & \frac{\sqrt{3}}{6}(a+b)+\frac{\sqrt{3}}{12}(1+c)
\end{array}
\right)\nonumber\\
&&\otimes \left(
\begin{array}{cccc}
\frac{3}{8} & \frac{b}{4}+\frac{1}{8} & \frac{b}{4}+\frac{1}{8} & \frac{b}{4}+\frac{1}{8}\\
\frac{a}{4}+\frac{1}{8} & \frac{1+3c}{12}+\frac{1}{24} & \frac{1}{8} & \frac{1}{8}\\
\frac{a}{4}+\frac{1}{8} & \frac{1}{8} & \frac{1}{8} & \frac{1+3c}{12}+\frac{1}{24} \\
\frac{a}{4}+\frac{1}{8} & \frac{1}{8} & \frac{1+3c}{12}+\frac{1}{24} & \frac{1}{8}
\end{array}
\right),\nonumber
\end{eqnarray}
we have $$\parallel\mathcal{P}_{\varrho^{'}}^{\underline{A}BC}\parallel_{tr}=\frac{3\sqrt{3}}{4}\sqrt{b^{2}+c^{2}+b+c+1}+\frac{3}{4}\sqrt{2+(b+c-1)^{2}-b-c}.$$
As shown in FIG \ref{fig:ex1abc}, its minimum value is greater than 1. Hence, $\varrho^{'}$ is entangled for any positive number $a, b, c$ satisfying $a+b+c=1$. When $a=b=c=\frac{1}{3}$, the state $\varrho^{'}$ is reduced to $\varrho$, which has been proved entangled.

\begin{figure}[h]
\centering
\includegraphics[width=9in]{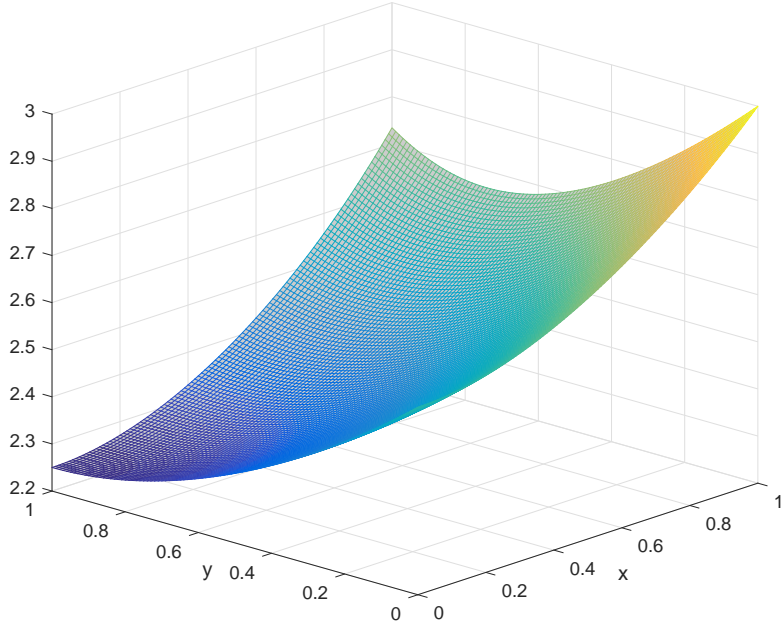}
\caption{The value of $\|\mathcal{P}_{s}^{AB|CD}\|_{tr}$ for $x\in[0, 1]$, $y\in[0, 1]$.}
\label{fig:ex1abcd}
\end{figure}

We construct a four-partite quantum state $\rho_{ABCD}$, which is similar to $\varrho^{'}$,  and  detect the entanglement of it by using SIC-POVMs.
\begin{eqnarray*}
\rho_{ABCD}&&=x|\psi ^{+}\rangle \langle\psi ^{+}|_{AB}\otimes |00\rangle \langle 00|_{CD}+y|\psi ^{+}\rangle \langle\psi ^{+}|_{AC}\otimes |00\rangle \langle 00|_{BD}\nonumber\\
&&+z|\psi ^{+}\rangle \langle\psi ^{+}|_{AD}\otimes |00\rangle \langle 00|_{BC},\nonumber
\end{eqnarray*}
with $x+y+z=1$, $x, y, z\geq 0$.
By directly calculation, we can get the trace norm of correlation matrix $\mathcal{P}_{s}^{AB|CD}$, that is
\begin{eqnarray*}
\|\mathcal{P}_{s}^{AB|CD}\|_{tr}&&=\frac{3}{16}(\sqrt{3(a^{2}+a+1)}+\sqrt{a^{2}-3a+3})\nonumber\\
&&\times(\sqrt{a^{2}+ab+2a+b^{2}+b+1}+\sqrt{a^{2}+3ab-6a+3b^{2}-9b+9}).
\end{eqnarray*}
The value of $\|\mathcal{P}_{s}^{AB|CD}\|_{tr}$ is greater than one, which is shown in FIG \ref{fig:ex1abcd}. Thus, according to criterion in Theorem \ref{Theorem1}, $\rho_{ABCD}$ is entangled for any positive number $x, y, z$ satisfying $x+y+z=1$.
\eex

\begin{figure}[h]
\centering
\includegraphics[width=8in]{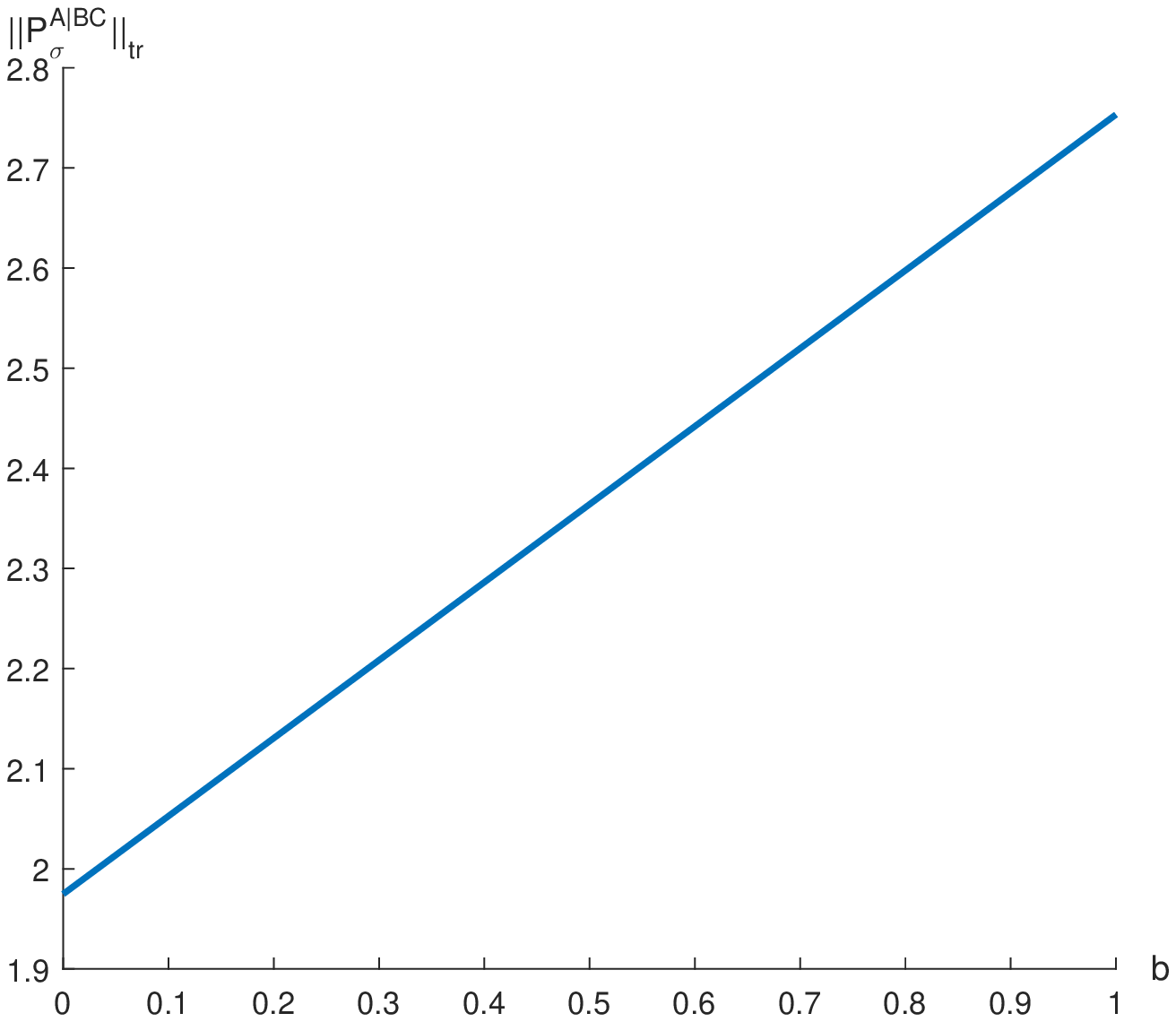}
\caption{The abscissa and ordinate represent $b$ and $\parallel\mathcal{P}_{\sigma}^{\underline{A}BC}\parallel_{tr}$, respectively.  For $0<b<1$, $\parallel\mathcal{P}_{\sigma}^{\underline{A}BC}\parallel_{tr}>1$.}
\label{fig:sigmab}
\end{figure}

\bex
\label{example3}
Consider the quantum state,
\begin{eqnarray}
\sigma=\frac{7b}{7b+1}\sigma_{insep}+\frac{1}{7b+1}|\phi_{b}\rangle\langle\phi_{b}|,
\end{eqnarray}
where $$\sigma_{insep}=\frac{2}{7}(|\psi_{1}\rangle\langle\psi_{1}|+|\psi_{2}\rangle\langle\psi_{2}|+|\psi_{3}\rangle\langle\psi_{3}|)+\frac{1}{7}|011\rangle\langle011|,$$
$$|\phi_{b}\rangle=|1\rangle\otimes (\sqrt{\frac{1+b}{2}}|00\rangle+\sqrt{\frac{1-b}{2}}|10\rangle),$$
$$|\psi_{1}\rangle=\frac{1}{\sqrt{2}}(|000\rangle+|101\rangle),$$
$$|\psi_{2}\rangle=\frac{1}{\sqrt{2}}(|001\rangle+|110\rangle),$$
$$|\psi_{3}\rangle=\frac{1}{\sqrt{2}}(|010\rangle+|111\rangle).$$
The state $\sigma$ is positive under arbitrary partial trnspositions for $0<b\leq1$. By using the SIC-POVMs in (\ref{e15}), we have $\parallel\mathcal{P}_{\sigma}^{\underline{A}BC}\parallel_{tr}>1$ for $0<b<1$, which is shown in FIG \ref{fig:sigmab}. So Theorem \ref{Theorem1} can detect the entangled state. But the criterion in \cite{YAMA} can only detect entanglement of $\sigma$ for $0< b\leq 0.10207$ and $0.38441< b \leq1$. Hence, our criterion is more efficient than the criterion in \cite{YAMA}.
\eex

\bex
\label{example2}
Consider the tripartite state \cite{CHB},
\begin{eqnarray}
\rho_{ABC}=\frac{1}{4}(I_{8}-\sum^{4}_{k=1}|\phi_{k}\rangle\langle\phi_{k}|),\nonumber
\end{eqnarray}
where
\begin{eqnarray}
&&|\phi_{1}\rangle=|0,1,+\rangle,|\phi_{2}\rangle=|1,+,0\rangle,|\phi_{3}\rangle=|+,0,1\rangle,\nonumber\\
&&|\phi_{4}\rangle=|-,-,-\rangle,|\pm\rangle=\frac{1}{\sqrt{2}}(|0\rangle\pm|1\rangle).\nonumber
\end{eqnarray}
It was shown in \cite{CHB} that the state is biseparable under any bipartite partitions $A|BC$, $B|AC$ and $C|AB$, but it is still entangled.
\begin{figure}[h]
\centering
\includegraphics[width=9in]{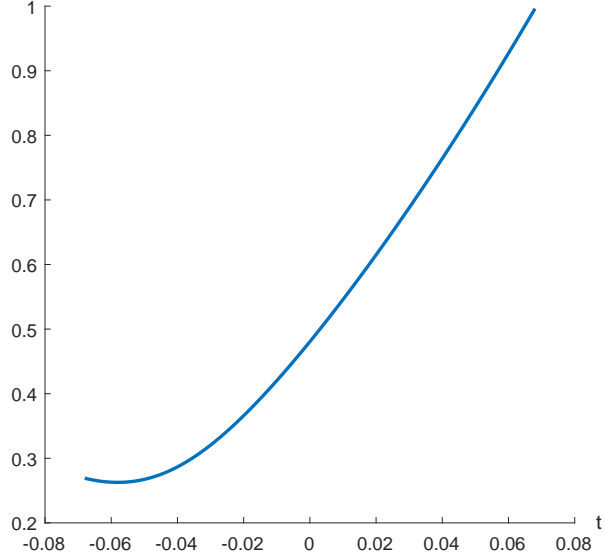}
\caption{The value of $\parallel\mathcal{P}_{\rho_{ABC}}^{\underline{A}BC}\parallel_{tr}-\frac{4a+1}{6}\sqrt{\frac{4a+1}{6}}$ for $t\in[-0.068, 0.068]$.}
\label{fig:ex2abc}
\end{figure}
By using the GSIC-POVMs in (\ref{GSIC1}), the correlation matrix is obtained,
\begin{eqnarray}
\mathcal{P}_{\rho_{ABC}}^{\underline{A}BC}&&=\left(
\begin{array}{cccc}
\frac{\sqrt{2}}{2}t+\frac{1}{4}& 0 & 0 & 0\\
0 & \frac{\sqrt{2}}{2}t+\frac{1}{4} & 0 & 0\\
0 & 0 & \frac{\sqrt{2}}{2}t+\frac{1}{4} & 0 \\
0 & 0 & 0 & \frac{3\sqrt{2}}{2}t+\frac{1}{4}
\end{array}
\right)\nonumber\\
&&\otimes \left(
\begin{array}{cccc}
(\frac{\sqrt{2}}{2}t+\frac{1}{4})^{2}-\frac{25}{8}t^{2} & (\frac{\sqrt{2}}{2}t+\frac{1}{4})^{2}-\frac{25}{8}t^{2} & \frac{\sqrt{2}}{4}t-\frac{21}{8}t^{2}+\frac{1}{16} & \frac{27}{8}t^{2}+\frac{\sqrt{2}}{2}t+\frac{1}{16}\\
(\frac{\sqrt{2}}{2}t+\frac{1}{4})^{2}+\frac{5}{8}t^{2} & (\frac{\sqrt{2}}{2}t+\frac{1}{4})^{2}-\frac{1}{8}t^{2} & \frac{9}{8}t^{2}+\frac{\sqrt{2}}{4}t+\frac{1}{16} & \frac{9}{8}t^{2}+\frac{\sqrt{2}}{2}t+\frac{1}{16}\\
\frac{39}{8}t^{2}+\frac{\sqrt{2}}{4}t+\frac{1}{16} & \frac{\sqrt{2}}{4}t-\frac{3}{8}t^{2}+\frac{1}{16} & \frac{39}{8}t^{2}+\frac{\sqrt{2}}{4}t+\frac{1}{16} & \frac{\sqrt{2}}{2}t-\frac{9}{8}t^{2}+\frac{1}{16} \\
\frac{27}{8}t^{2}+\frac{\sqrt{2}}{2}t+\frac{1}{16} & \frac{9}{8}t^{2}+\frac{\sqrt{2}}{2}t+\frac{1}{16} & \frac{27}{8}t^{2}+\frac{\sqrt{2}}{2}t+\frac{1}{16} & (\frac{3\sqrt{2}}{2}t+\frac{1}{4})^{2}-\frac{9}{8}t^{2}
\end{array}
\right),\nonumber
\end{eqnarray}
 then we have $\parallel\mathcal{G}_{\rho_{ABC}}^{\underline{A}BC}\parallel_{tr}-\frac{4a+1}{6}\sqrt{\frac{4a+1}{6}}>0$. We describe it in FIG \ref{fig:ex2abc}. So Theorem \ref{Theorem2} can detect the entangled state.
\eex

\bex
\begin{figure}[h]
\centering
\includegraphics[width=9in]{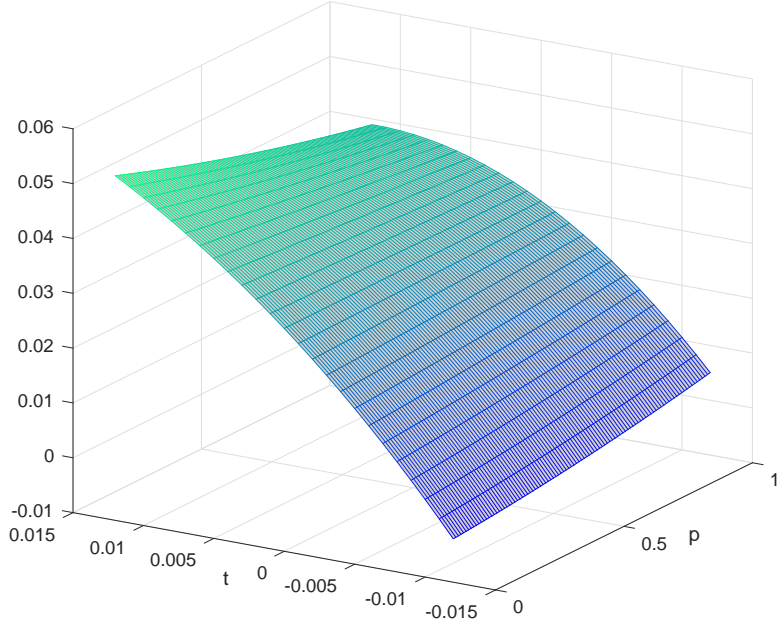}
\caption{The value of $\parallel\mathcal{G}_{\rho}^{\underline{A}BC}\parallel_{tr}-\sqrt{\frac{9a_{1}+1}{12}}\sqrt{\frac{9a_{2}+1}{12}}\sqrt{\frac{4a_{3}+1}{6}}$ for $t\in[-0.012, 0.012]$.}
\label{fig:lastex}
\end{figure}
\label{example4}
Consider the quantum state mixed with the white noise,
\begin{eqnarray}
\rho=\frac{1-p}{18}I+p|\varphi\rangle\langle\varphi|,\nonumber
\end{eqnarray}
where $|\varphi\rangle=\frac{1}{\sqrt{5}}[(|10\rangle+|21\rangle)|0\rangle+(|00\rangle+|11\rangle+|22\rangle)|1\rangle]$ is an entangled state \cite{MFC}, and $0\leq p\leq 1.$

By using the GSIC-POVMs in (\ref{GSIC1}) and (\ref{GSIC2}), we can get the value of $\parallel\mathcal{G}_{\rho}^{\underline{A}BC}\parallel_{tr}-\sqrt{\frac{9a_{1}+1}{12}}\sqrt{\frac{9a_{2}+1}{12}}\sqrt{\frac{4a_{3}+1}{6}}$. Here the $a_{i}$, $i=1, 2, 3$ are the parameters of subsystem.  Theorem \ref{Theorem2} can detect the entanglement of $\rho$ for $0.35\leq p\leq 1$. However, the criterion in \cite{YAMA} can only detect the entanglement for $p=1$, hence our criterion is better than that in \cite{YAMA}. We describe it in FIG \ref{fig:lastex}.
\eex

\section{CONCLUSION}
\label{conclu}
The detection of entanglement is a basic problem in quantum theory. We have studied the entanglement criteria based on SIC measurement and GSIC measurement. We have obtained effective criteria to detecting entanglement for tripartite system, which could be extended to multipartite system. Comparing with the existing criterion, the criterion can detect entangled states by theoretical analysis and numerical examples. The method used in this paper can also be generalized to arbitrary multipartite qudit systems. It would be also worthwhile to investigate the multipartite genuinely entanglement.

\bigskip
\noindent{\bf Acknowledgments}\, Authors were supported by the NNSF of China (Grant No. 11871089), and the Fundamental
Research Funds for the Central Universities (Grant Nos. KG12080401 and ZG216S1902).

\end{document}